\documentclass[12pt]{article}
\usepackage{amsfonts}
\usepackage{amsmath}
\usepackage{amssymb}
\usepackage{amscd}
\begin{document}
\title{\bf General properties of the evolution\\
of unstable states at long times}
\author{K. Urbanowski\footnote{email: K.Urbanowski@proton.if.uz.zgora.pl}\\
University of Zielona Gora, Institute of Physics,\\ ul. Prof. Z. Szafrana 4a,
65--516 Zielona Gora, Poland.}
\maketitle
\begin{abstract}
An effect generated by the nonexponential behavior
of the survival amplitude of an unstable state at the long time region is considered.
It is known that this  amplitude
tends to zero as $t$ goes to the infinity more slowly than any exponential function of $t$.
Using methods of asymptotic analysis
we find the asymptotic form of this amplitude  in the long time region in a
general model independent case.
We find that the long time behavior  of this amplitude affects the form of
the instantaneous energy of  unstable states: This energy should be much smaller for suitably
long times $t$ than the energy of this state for $t$ of the order of the lifetime of
the considered unstable state.
\end{abstract}

\section{Introduction}

Unstable states $|u\rangle$
of the system under considerations are characterized by their decay law,
${\cal P}_{u}(t)$,
\begin{equation}
{\cal P}_{u}(t) = |a(t)|^{2}, \label{P(t)}
\end{equation}
where
\begin{equation}
a(t) = \langle u|u (t) \rangle,  \label{a(t)}
\end{equation}
is the  probability amplitude of finding the system at the
time $t$ in the initial state $|u\rangle \in {\cal H}$ prepared at time $t_{0}
= 0$,  ${\cal H}$ is the Hilbert space of states of the system,
 $\| \,|u\rangle \,\|=1$ so $a(0) = 1$,
and $|u(t)\rangle \in {\cal H}$ solves the Sch\"{o}dinger equation
\begin{equation}
i\hbar \frac{\partial}{\partial t} |u (t) \rangle = H |u
(t)\rangle, \;\;\;\;\; |u(0) \rangle = |u\rangle, \label
{Schrod}
\end{equation}
where $H$ denotes the
total selfadjoint Hamiltonian for the system.
From basic principles of quantum theory it is known that the nondecay
amplitude $a(t)$, and thus the decay law ${\cal P}_{u}(t)$ of the
unstable state $|u\rangle$ decaying in the vacuum, are completely determined by the
density of the energy distribution $\omega(\varepsilon)$ for the system
in this state \cite{Fock},
\begin{equation}
a(t) = \int_{-\infty}^{+\infty} \omega(\varepsilon)\;
e^{\textstyle{-\frac{i}{\hbar}\,\varepsilon\,t}}\,d\varepsilon.
\label{a-spec}
\end{equation}
where $\omega(\varepsilon) \geq 0$.

Note that (\ref{a-spec}) and (\ref{a(t)}) together with the condition that $|u\rangle$ is a normalized vector
mean that there must be
\begin{equation}
a(0) = \int_{-\infty}^{+\infty} \omega (\varepsilon)\,d\varepsilon = 1.
\label{a(0)-spec}
\end{equation}
From the last property  one concludes that $\omega(\varepsilon)$ is an
absolutely integrable function, $\omega(\varepsilon) \in L_{1}(-\infty, \infty)$.
So the amplitude $a(t)$ is the Fourier
transform of $\omega (\varepsilon)$ (see (\ref{a-spec})) and thus
from the Riemann--Lebesgue Lemma it
follows that $a(t)$ must tend to
zero as $t \rightarrow \infty$ \cite{Fock,kolgomorov}.

A condition, which is necessary on physical grounds, that $H$
has a spectrum bounded from below, $ Spec. (H) = [E_{min}, +\infty)$,
and $E_{min} > - \infty$, reduces the set of functions $\omega(\varepsilon) \in L_{1}(-\infty, \infty)$
to such $\omega(\varepsilon) \in L_{1}(-\infty, \infty)$ that
$\omega(\varepsilon) = 0$ for $\varepsilon < E_{min}$ and
$\omega(\varepsilon) \geq 0$ for $\varepsilon \geq E_{min}$. Thus in fact the integration in
(\ref{a-spec}), (\ref{a(0)-spec}) and in similar formulae is taken over $\varepsilon \in [E_{min}, +\infty)$.

In \cite{Khalfin} assuming that the spectrum of $H$ must be bounded
from below and using the Paley--Wiener
Theorem \cite{Paley} it was proved that in the case of unstable
states there must be
\begin{equation}
|a(t)| \; \geq \; A\,e^{\textstyle - b \,t^{q}}, \label{|a(t)|-as}
\end{equation}
for $|t| \rightarrow \infty$. Here $A > 0,\,b> 0$ and $ 0 < q < 1$.
This means that the decay law ${\cal P}_{u}(t)$ of unstable
states decaying in the vacuum, (\ref{P(t)}), can not be described by
an exponential function of time $t$ if time $t$ is suitably long, $t
\rightarrow \infty$, and that for these lengths of time ${\cal
P}_{u}(t)$ tends to zero as $t \rightarrow \infty$  more slowly
than any exponential function of $t$. The analysis of the models of
the decay processes shows that ${\cal P}_{u}(t) \simeq
e^{\textstyle{- \frac{\gamma_{u}^{0} t}{\hbar}}}$, (where
$\gamma_{u}^{0}$ is the decay rate of the state $|u \rangle$),
to a very high accuracy  for a wide time range $t$: From $t$
suitably greater than some $T_{0} \simeq t_{0}= 0$ but $T_{0} >
t_{0}$ (${\cal P}_{u}(t)$ has nonexponential power--like form
for short times $t \in (t_{0},T_{0})$ -- see, e.g. \cite{Khalfin,Fonda,Peres})
up to $t \gg \tau_{u} = \frac{\hbar}{\gamma_{u}^{0}}$
and smaller than $t = t_{as}$, where $t_{as}$ denotes the
time $t$ for which the long time nonexponential deviations of $a(t)$
begin to dominate (see eg., \cite{Khalfin}, \cite{Fonda} --
\cite{Sluis}). From this analysis it follows that in the general
case the decay law ${\cal P}_{u}(t)$ takes the inverse
power--like form $t^{- \lambda}$, (where $\lambda
> 0$), for suitably large $t \geq t_{as}\gg \tau_{u}$
\cite{Khalfin}, \cite{Fonda} -- \cite{Goldberger}. This effect is in
agreement with the  general result (\ref{|a(t)|-as}).  Effects of this type
are sometimes called the "Khalfin effect" (see eg.
\cite{Arbo}).

The problem how to detect possible deviations from the exponential form of
${\cal P}_{u}(t)$ in the long time region  has been attracting attention of physicists since the
first theoretical predictions of such an effect \cite{Wessner,Norman1,Greenland}.
The tests that have been performed over many years to examine the form of the decay laws
for  $t \gg \tau_{u}$  have not indicated any deviations from the exponential form of ${\cal P}_{u}(t)$ in the
long time region.
Nevertheless, conditions leading to the nonexponetial behavior
of the amplitude $a(t)$ at long times were studied theoretically \cite{seke} -- \cite{jiitoh}.
Conclusions following from these studies were applied successfully in experiment described  in \cite{rothe},
where the experimental evidence of deviations from the exponential decay law at long times was
reported. This result gives rise to another problem which now becomes important:
if and how the long time deviations from the exponential decay law depend on the model considered
(that is, on the form of $\omega(\varepsilon)$), and
if (and how) these deviations  affect the  energy of the unstable state
and its decay rate in the long time  region.

\section{General long time properties of the nondecay amplitude}

Many fundamental and general model independent properties of the nondecay amplitude $a(t)$
follow from the fact that $a(t)$ is the  Fourier transform of an absolutely integrable
function $\omega (\varepsilon)$. So,   if one assumes  that all derivatives $\omega^{(k)} (\varepsilon)$,
($k=0,1, \ldots , n$), exist and
that $\omega^{(k)} (\varepsilon) \in L_{1}(-\infty, \infty)$ for all these $k$,
(where $\omega^{(0)}(\varepsilon) = \omega (\varepsilon)$), then analyzing general properties of the
Fourier transforms it is easy to find that
for $t \rightarrow \infty$
\begin{equation}
|a(t)| \,\leq \, \frac{C}{t^{n}}, \label{a<infty}
\end{equation}
where $0 < C < \infty$ (see \cite{kolgomorov}). So if the derivative of $\omega (\varepsilon)$ exists and
$\omega^{(1)} (\varepsilon) \in L_{1}(-\infty, \infty)$ then taking into account relation
(\ref{|a(t)|-as}) the following estimation follows
\begin{equation}
 \frac{C}{t} \geq |a(t)| \geq A\,e^{\textstyle - b \,t^{q}}
\label{>a>}
\end{equation}
as $t \rightarrow \infty$.

Much more information about asymptotic properties of $a(t)$ being the Fourier
transform of $\omega (\varepsilon)$  one can find if  the assumption required by physics
that the spectrum of $H$ is bounded from below is used, eg. by $\varepsilon = E_{min} > -\infty$. This means
that from this time
we will consider only
such $\omega({\varepsilon)} \in L_{1}(-\infty, \infty)$ that
$\omega({\varepsilon)} = 0$ for $\varepsilon < E_{min}$ and
$\omega({\varepsilon)} \geq 0$ for $\varepsilon > E_{min}$. In this case there is
\begin{equation}
a(t) = \int_{E_{min}}^{+\infty} \omega(\varepsilon)\;
e^{\textstyle{-\frac{i}{\hbar}\,\varepsilon\,t}}\,d\varepsilon.
\label{a-spec>E-min}
\end{equation}
instead of the formula (\ref{a-spec}) for the amplitude $a(t)$. Using methods
of the asymptotic analysis it is not
difficult to find an asymptotic expansion for large values of  $t$ of the
Fourier integral of this type \cite{olver,erdelyi,copson}.

Let us consider for a start   relatively simple case when $\lim_{\varepsilon \rightarrow E_{min}+}
\;\omega (\varepsilon)\stackrel{\rm def}{=} \omega_{0}>0$.   Let  derivatives  $\omega^{(k)}(\varepsilon)$, ($k= 0,1,2, \ldots, n$),  be continuous
in %\linebreak
$[E_{min}, \infty)$, (that is let for $\varepsilon > E_{min}$ all
$\omega^{(k)}(\varepsilon)$ be continuous
and all the limits \linebreak
$\lim_{\varepsilon \rightarrow E_{min}+}\,\omega^{(k)}(\varepsilon)$ exist)  and
let all these $\omega^{(k)}(\varepsilon)$ be absolutely integrable functions then,
as  can be easily verified,  (see \cite{olver,erdelyi,copson}),
\begin{equation}
a(t) \; \begin{array}{c}
          {} \\
          \sim \\
          \scriptstyle{t \rightarrow \infty}
        \end{array}
        \;- \frac{i\hbar}{t}\;e^{\textstyle{-\frac{i}{\hbar}E_{min} t}}\;
        \sum_{k = 0}^{n-1}(-1)^{k} \,\big(\frac{i\hbar}{t}\big)^{k}\,\omega^{(k)}_{0},
        \label{a-omega}
\end{equation}
 where $\omega^{(k)}_{0}  \stackrel{\rm def}{=} \lim_{\varepsilon \rightarrow E_{min}+}
\;\omega^{(k)} (\varepsilon)$.

Let us now consider a more complicated form of the density $\omega (\varepsilon)$. Namely
let $\omega (\varepsilon)$ be of the form
\begin{equation}
\omega (\varepsilon) = ( \varepsilon - E_{min})^{\lambda}\;\eta (\varepsilon)\; \in \; L_{1}(-\infty, \infty),
\label{omega-eta}
\end{equation}
where $0 < \lambda < 1$ and it is assumed that $\eta^{(k)}(\varepsilon)$,
($k= 0,1,2, \ldots, n$), exist and they are continuous
in
$[E_{min}, \infty)$, and  limits
$\lim_{\varepsilon \rightarrow E_{min}+}\;\eta^{(k)}(\varepsilon)$ exist,
$\lim_{\varepsilon \rightarrow \infty}\;( \varepsilon - E_{min})^{\lambda}\,\eta^{(k)}(\varepsilon) = 0$
for all above mentioned $k$, then
\begin{eqnarray}
a(t) & \begin{array}{c}
          {} \\
          \sim \\
          \scriptstyle{t \rightarrow \infty}
        \end{array} &
        - \frac{i\hbar}{t}\;\lambda\;e^{\textstyle{-\frac{i}{\hbar}E_{min} t}}\;
        \Big[\alpha_{n}(t) + \big(-\,\frac{i\hbar}{t}\,\big)\,\alpha_{n-1}(t) \nonumber \\
        && + \big(-\,\frac{i\hbar}{t}\big)^{2}\,\alpha_{n-2}(t)\nonumber \\&& +
        \big(-\,\frac{i\hbar}{t}\big)^{3}\,\alpha_{n-3}(t)+ \dots \Big],
        \label{a-eta}
\end{eqnarray}
where (compare \cite{erdelyi,copson})
\begin{equation}
\alpha_{n-k}(t) = \sum_{l=0}^{n-k-1}\,\frac{\Gamma (l + \lambda)}{l!}\;\,e^{\textstyle{-\,i\,\frac{\pi (l + \lambda +2)}{2}}}
\;\eta_{0}^{(l + k)}\,\big(\frac{\hbar}{t} \big)^{l + \lambda}, \label{alpha}
\end{equation}
and $\eta^{(j)}_{0}  = \lim_{\varepsilon \rightarrow E_{min}+}
\;\eta^{(j)} (\varepsilon)$, $\eta^{(0)}(\varepsilon) = \eta (\varepsilon)$ and $j = 0,1, \ldots,n$.

The main difference between asymptotic expansions (\ref{a-omega}) and (\ref{a-eta}) is that the  amplitude $a(t)$ in (\ref{a-omega}) is obtained as the Fourier transform (\ref{a-spec>E-min}) of such $\omega(\varepsilon)$ that $\omega(\varepsilon) = 0$ for $\varepsilon < E_{min}$ and $\omega(E_{min}) >0$ whereas the expansion (\ref{a-eta})  is the asymptotic expansion of the Fourier transform  (\ref{a-spec>E-min}) for another type $\omega (\varepsilon)$:  namely for $\omega (\varepsilon)$ such that
$\omega(E_{min}) = 0$ (see (\ref{omega-eta})).

From (\ref{a-omega}), (\ref{a-eta}) it follows that the estimation (\ref{>a>}) is true for
all physically admissible $\omega(\varepsilon)$. This estimation holds for $t \rightarrow \infty$,
strictly speaking for $t > t_{as}$. For $t \sim \tau_{u}$, $t < t_{as}$ the nondecay amplitude $a(t)$ takes the form
\begin{equation}
a(t) \;\simeq\; e^{\textstyle{-i\frac{t}{\hbar}(E_{u}^{0} - \frac{i}{2}\,\gamma_{u}^{0})}},
\label{a(t)-sim}
\end{equation}
to a very high accuracy \cite{Khalfin}, \cite{Fonda} -- \cite{Arbo}, \cite{seke} -- \cite{jiitoh}.
In this formula $E_{u}^{0}$ denotes the measured energy of the unstable particle described by the
state--vector $|u\rangle$. There is $E_{u}^{0}\,>\,E_{min}$.

For the energy densities $\omega(\varepsilon)$ leading to the asymptotic form of the amplitude $a(t)$ of type (\ref{a-omega}) the time $t_{as}$ can be found
by comparing the square of the modulus of the amplitude $a(t)$
from the relation (\ref{a(t)-sim}) and the square of the modulus of the leading component of (\ref{a-omega}).
So $t_{as}$ can be found by solving the following transcendental equation
\begin{equation}
e^{\textstyle{-\,\frac{\gamma_{u}^{0}}{\hbar}\,t}}\;=\;\hbar^{2}\;
\Big(\,\frac{\omega (E_{min})}{t}\,\Big)^{2}.
\end{equation}
This means that the value of $t_{as}$ depends on the model considered: it depends on
the density $\omega(\varepsilon)$ and on the $\gamma_{u}^{0}$. If $\omega(\varepsilon)$ has the form (\ref{omega-eta}) then a similar method can be used to find a corresponding equation for $t_{as}$.

\section{Energy of unstable states at long time region}

Using the Khalfin's estimation (\ref{|a(t)|-as}) of the decay law (\ref{P(t)}) one can examine
the asymptotic properties of the decay rate $\gamma_{u}$ of an unstable state $|u\rangle$.
In a general case the decay rate $\gamma_{u}$ equals,
\begin{equation}
\gamma_{u}\, =\,\gamma_{u}(t)\,\stackrel{\rm def}{=}\, -
\;\frac{\hbar}{{\cal P}_{u}(t)}\;\frac{\partial {\cal
P}_{u}(t)}{\partial t}. \label{gamma-1}
\end{equation}
From (\ref{gamma-1}), (\ref{P(t)}) and (\ref{a(t)-sim}) one infers that %for $t < t_{as}$
\begin{equation}
\gamma_{u}\, =\,\gamma_{u}(t) \equiv \gamma_{u}^{0}, \;\;\;\;\;\;(\text{for}\;t < t_{as}),
\label{gamma-0}
\end{equation}
which is obvious.
From (\ref{|a(t)|-as}) it follows that in the asymptotic case $t \rightarrow \infty$
the decay rate $\gamma_{u}(t)$ can not be larger than \cite{PRA},
\begin{equation}
\gamma_{u}(t)\;\begin{array}{c}
                    {} \\
                     \sim \\
                    \scriptstyle{t \rightarrow \infty}
                  \end{array}
\;bq\;t^{-\mu}, \label{gamma-as}
\end{equation}
where $\mu \equiv 1 - q > 0$. So for $t > t_{as}$ one finds that
${\gamma_{u}(t)\vline}_{\;t \rightarrow \infty} < \gamma_{u}^{0}$ for every physically admissible $\omega (\varepsilon)$,
and in general that
$\lim_{t \rightarrow \infty} \gamma_{u}(t)\;=\;0$.

The problem is how the energy $E_{u}$ of the
unstable state $|u\rangle$ behaves for $t > t_{as}$. The solution of this problem follows from
the observation that the amplitude $a(t)$ can be found either by solving the Schr\"{o}dinger equation (\ref{Schrod}) or
using the equation for the projection of the state vector ( see \cite{PRA} and references one can find therein),
which in the case of one--dimensional subspace  ${\cal H}_{||}$ of
states  ${\cal H}$  spanned by the normalized vector
$|u\rangle$  has the following simple form
\begin{equation}
i \hbar\, \frac{\partial a(t)}{\partial t} \;=\;
h_{u}\,\;a(t),\;\;\; a(0) =1, \label{eq-for-h}
\end{equation}
where $h_{u}$ is  the "effective Hamiltonian" for the
one--dimensional subspace of states ${\cal H}_{||}$.
In general, $h_{u}$ can depend on time $t$, $h_{u}\equiv
h_{u}(t)$ \cite{PRA,horwitz}. One meets this effective Hamiltonian when one starts
with the Schr\"{o}dinger Equation (\ref{Schrod}) for the total state
space ${\cal H}$ and looks for the rigorous evolution equation for
the distinguished subspace of states ${\cal H}_{||} \subset {\cal
H}$. There are many approximate methods to calculate  $h_{u}$ \cite{PRA}
but taking into account the problem raised above
their use is not necessary. It is sufficient to use the property
that the exact effective Hamiltonian $h_{u}(t)$ must
fulfill the following identity \cite{PRA}
\begin{equation}
h_{u}\,\equiv \,h_{u}(t)\, \stackrel{\rm def}{=}\,  i \hbar\,
\frac{\partial a(t)}{\partial t} \; \frac{1}{a(t)}\,.
\label{h}
\end{equation}
Direct application (\ref{h}) to the relation (\ref{a(t)-sim}) yields
\begin{equation}
h_{u}(t) = h_{u}^{0} \equiv E_{u}^{0} - \frac{i}{2}\,\gamma_{u}^{0},\;\;\;\;\;(\text{for}\; \,t < t_{as}),
\label{h-0}
\end{equation}
which could be expected. Note that from (\ref{gamma-1}) and (\ref{h}) it follows that simply
\begin{equation}
\gamma_{u}(t)\,=\,-\,2\,\Im\,(h_{u}(t)).\label{G(t)}
\end{equation}
Similarly, the real part of $h_{u}(t)$ is the instantaneous energy, $E_{u}(t)$, of the
system in the state $|u\rangle$ under
considerations
\begin{eqnarray}
E_{u}&\equiv& E_{u}(t) = \Re\,(h_{u}(t)).
\label{E(t)}
\end{eqnarray}
(Here $\Re\,(z)$ and $\Im\,(z)$ denote the real and imaginary parts
of $z$ respectively).

Note that relations (\ref{h}) and (\ref{eq-for-h}) establish a direct
connection between the amplitude $a(t)$ for the state $|u
\rangle$ and the exact effective Hamiltonian $h_{u}(t)$ governing
the time evolution in the one--dimensional subspace ${\cal H}_{\|}
\ni |u\rangle$. Thus the use of  the relation (\ref{h}) is one of the most
effective tools for the accurate analysis of the early-- as well as
the long--time properties of the instantaneous energy and decay rate for a given
qausistationary state $|u (t) \rangle$.

Now let us analyze  the asymptotic properties of $h_{u}(t)$ for $t \rightarrow \infty$. For the densities $\omega(\varepsilon)$ leading to %Using
asymptotic expansion (\ref{a-omega})
one finds that for $t\rightarrow \infty$,
\begin{eqnarray}
i\hbar\,\frac{\partial a(t)}{\partial t}\; & \begin{array}{c}
                                               {} \\
                                               \simeq\\
                                               \scriptstyle{t \rightarrow \infty}
                                             \end{array} &
 \; E_{min}\,a(t)\;+\;\big(\,\frac{i\hbar}{t}\,\big)^{2}\,\;
e^{\textstyle{-\,\frac{i}{\hbar}\,E_{min}\,t}}\;\times \nonumber \\&& \times \;\Big\{\,\omega_{0}\;
-\; 2\,\omega^{(1)}_{0}\,\big(\frac{i\hbar}{t}\big)\;
 + \;3\;\omega_{0}^{(2)}\;\big(\frac{i\hbar}{t}\big)^{2} \nonumber \\&&
-\;4\;\omega_{0}^{(3)}\;\big(\frac{i\hbar}{t}\big)^{3}\;+
\ldots\Big\}.
\label{da-infty}
\end{eqnarray}
The next step is to use the relation (\ref{h}). So one should now divide (\ref{da-infty}) by (\ref{a-omega}) and then
collect together all components  of the same order with respect to $(\,\frac{\hbar}{t}\,)$. As the result
one obtains the asymptotic form of $h_{u}(t)$ for $t \rightarrow \infty$,
\begin{eqnarray}
h_{u}^{\infty}(t)\; \stackrel{\rm def}{=}\;
         {h_{u}(t)\,\vline}_{\;t \rightarrow \infty} \; &=& \;E_{min} \,-\,i\,\frac{\hbar}{t}\nonumber \\
         &&-\frac{\omega_{0}^{(1)}}{\omega_{0}}\;\big(\,\frac{\hbar}{t}\,\big)^{2}\;+\;\ldots\;\, . \label{h-infty}
\end{eqnarray}
One obtains a similar form of $h_{u}^{\infty}(t)$ for the amplitude $a(t)$ given by formulae (\ref{a-eta}), (\ref{alpha}). Indeed starting from
(\ref{a-eta}) one finds after some algebra that
\begin{equation}
h_{u}^{\infty}(t) \;=
\;
E_{min}\,-\,c_{1}\,
\frac{\hbar}{t}\, -\,c_{2}
\,\big(\frac{\hbar}{t}\big)^{2}\,-\,c_{3}\,\big(\frac{\hbar}{t}\big)^{3}\,+\,\ldots\,\,\, ,
\label{h-sim-w}
\end{equation}
where $c_{1}, c_{2}, c_{3}, \ldots$ are complex numbers with negative or positive real and imaginary parts.

A surprising conclusion following from the result (\ref{h-infty}) is that  in the long time region
the leading component of the asymptotic form of the decay rate $\gamma_{u}(t)$ has the same form for a large class of
physically admissible models,
\begin{equation}
\gamma_{u}^{\infty}(t)\; \stackrel{\rm def}{=}\;
         {\gamma_{u}(t)\,\vline}_{\;t \rightarrow \infty} \;
         \equiv \; -\,2\,\Im\,(h_{u}^{\infty}(t))\;\simeq \;2\;\frac{\hbar}{t}\;.
\label{G(t)-infty}
\end{equation}
Using the relation (\ref{h-infty}) we find that  in the long time region the instantaneous energy $E_{u}(t)$ takes the following form
\begin{eqnarray}
E_{u}^{\infty}(t)\; &\stackrel{\rm def}{=}&\;
         {E_{u}(t)\,\vline}_{\;t \rightarrow \infty} \;=\;\Re\,(h_{u}^{\infty}(t))\;\makebox[25mm]{} \nonumber \\
         &\simeq &\; E_{min}\;-\;\frac{\omega_{0}^{(1)}}{\omega_{0}}\;\big(\,\frac{\hbar}{t}\,\big)^{2}\;+\;\ldots\;\neq\;E_{u}^{0}.
         \label{E-infty}
\end{eqnarray}
As one can see
for all densities $\omega(\varepsilon)$ such that $\omega_{0} \equiv \omega(E_{min}) > 0$
the long time properties of the leading components of the energy $E_{u}(t)$, contrary to the properties of the decay rate $\gamma_{u}(t)$,
depend on the density $\omega (\varepsilon)$.
The result (\ref{E-infty}) seems to be even much more surprising than (\ref{G(t)-infty}).

We have
\begin{equation}
\lim_{t \rightarrow \infty}\;E_{u}^{\infty}(t)\;=\;E_{min}. \label{lim-E}
\end{equation}
Note that the same result follows from (\ref{h-sim-w}) which means that the relation (\ref{lim-E}) is a model independent.
Taking into account that $E_{u}^{0}\,>\,E_{min}$ the following conclusion follows: for every model
(that is for every $\omega (\varepsilon)$ ) there exists such $t_{\infty} \geq t_{as}$ that
\begin{equation}
E_{u}^{\infty}(t)\;<\;E_{u}^{0}, \;\;\;\;(\text{for}\;\;t\,>\,t_{\infty}). \label{E-infty<E-0}
\end{equation}
Note that results (\ref{h-infty}) -- (\ref{E-infty<E-0}) are purely quantum effects and that they follow
from basic assumptions of quantum theory.

\section{Final remarks}

The problem if the long time deviations from the exponential form of the decay law affect the energy of the decaying state
has been studied in \cite{epj-c-2008} using a model defined by $\omega(\varepsilon)$ having a form of the truncated Lorentz function and assuming that $E_{min}=0$. It is easy to verify that inserting into (\ref{a-omega}), (\ref{h-infty})  $E_{min}=0$ and $\omega(\varepsilon)$ used in \cite{epj-c-2008} reproduces relations obtained there. Results obtained in Sec. 2 and Sec. 3 show that long time behavior of the decay law ${\cal P}_{u}(t)$ as well as the effective Hamiltonian $h_{u}(t)$ do not depend on a specific form of the density $\omega(\varepsilon)$ but they depend rather on general  integral and analytic properties of the density $\omega(\varepsilon)$. From these results it follows that for all $\omega(\varepsilon)$ having the same integral and analytic properties the amplitudes $a(t)$ and the effective Hamiltonians $h_{u}(t)$ have the same long time behavior. So these results generalize and complete essentially analysis performed in \cite{epj-c-2008}.

The estimation (\ref{>a>}) of $|a(t)|$ for $t \rightarrow \infty$ follows
from basic assumption of quantum theory. Similarly, estimations (\ref{a-omega}), (\ref{a-eta}) of $a(t)$ at long time region
are obtained using only very general assumptions on the form of the energy density $\omega(\varepsilon)$ and the fundamental assumption that
there exists a minimal energy $E_{min} > - \infty $ in the system under considerations (i.e. that $Spec.(H) = [E_{min}, + \infty)$).
So, they should hold for every physical system fulfilling these general requirements. In general, as it follows from the analysis performed in Sec. 2, two types of the long time asymptotic expansion of the amplitude $a(t)$ can be observed depending on the continuity properties of the density $\omega(\varepsilon)$ at the point $\varepsilon = E_{min}$. If the density $\omega(\varepsilon)$ is a discontinuous function of $\varepsilon$ at $\varepsilon = E_{min}$: $\omega(\varepsilon < E_{min}) = 0$ and $\omega(E_{min}) >0$, then the long time asymptotic of $a(t)$ is given by the formula (\ref{a-omega}). A particular, typical example of such   $\omega(\varepsilon)$ is the truncated Lorentzian distribution function $\omega_{L}(\varepsilon)$,
\begin{equation}
\omega_{L}(\varepsilon) =
\frac{N}{2\pi}\,  {\it\Theta} (\varepsilon - E_{min}) \
\frac{\gamma_{u}^{0}}{(\varepsilon - E_{u}^{0})^{2} +
(\frac{\gamma_{u}^{0}}{2})^{2}}, \label{omega-BW}
\end{equation}
where $N$ is a normalization constant, $E_{u}^{0} > E_{min}$, and ${\it\Theta} (\varepsilon) = \{ 1\;\;{\rm for}\;\; \varepsilon \geq 0,\;\;{\rm and}\;\; 0 \;\;{\rm for}\;\;\varepsilon < 0\}$.    This distribution is the basis of many studies of decaying systems (see, eg. \cite{Khalfin,Sluis}).
Many unstable systems will have an initial state energy distribution $\omega(\varepsilon)$ that is close to  $\omega_{L}(\varepsilon)$ for all $\varepsilon$ values. It has been proved \cite{Fonda} that in such  cases the decay law ${\cal P}_{u}(t)$ for the system and decay law ${\cal P}_{u}^{L}(t)$ resulting from $\omega_{L}(\varepsilon)$ must be close to each other for all values of $t$.
In general the density $\omega (\varepsilon)$ having Lorentz (Breit--Wigner) shape is known from the response of a harmonically bound elektron with a dissipative term, models of resonance behavior  and many other physical problems.
On the other hand, if $\omega(\varepsilon)$ has the form (\ref{omega-eta}), that is if it is continuous at $\varepsilon = E_{min}$, then the long time form of the amplitude $a(t)$ is given by the relation  (\ref{a-eta}). A particular case of this type density distribution is $\omega(\varepsilon)$
which can be found when one considers short--range potential models of quasi--stationary states: One can find such a density for finite--width barriers as well as delta barriers and with or without a potential inside the barier, etc. (see, eg., \cite{parrot,lawrence,joichi,santra,winter,jiitoh} and references one can find therein). In general in the models mentioned
densities $\omega (\varepsilon )$ are proportional near $E_{min}$ to the square root of the energy $\varepsilon$,
\begin{equation}
\omega(\varepsilon) \sim \sqrt{(\varepsilon - E_{min})}, \label{s-r}
\end{equation}
for $\varepsilon \geq E_{min}$, (i.e. they correspond with $\lambda = \frac{1}{2}$  in (\ref{omega-eta}) and (\ref{a-eta}) ), and usually there is $E_{min} = 0$ in these models.
The another example of the density of this type is the density $\omega (\varepsilon)$ obtained when one considers the decay of an unstable particle into two particles \cite{Goldberger}.
The form of long time asymptotic expansions for $a(t)$ can differ from expansions (\ref{a-omega}) and (\ref{a-eta}) for $\omega(\varepsilon)$ being discontinuous at a point (or some points $j=1,2,\dots, $) $\varepsilon = E_{j} > E_{min}$.

The source of the effect described by
the relation (\ref{h-infty}) is the long time behavior of the amplitude $a(t)$. The relation (\ref{h}) establishes a direct connection between the
properties of the amplitude $a(t)$ and the properties of the instantaneous energy $E_{u}(t)$ and decay rate $\gamma_{u}(t)$ of the unstable $|u\rangle$ at the instant $t$, (see (\ref{h-0}), (\ref{E(t)}), (\ref{G(t)})). A possibility  observe the ``loss of energy'' described by relations (\ref{E-infty}), (\ref{E-infty<E-0}) may arise while trying to test the long time properties of the nondecay amplitude $a(t)$ after a suitable modification of such tests. So, considering a possibility of a suitable modification of the test described in \cite{rothe} in such a way that the emitted energy (frequency) of the luminescence decays could be measured which could make it possible to test relations (\ref{E-infty}), (\ref{E-infty<E-0}) seems worthwile.
In general, all these long time properties of unstable states  should not be expected to have an effect on laboratory processes
but it seems that they can affect some long time astrophysical processes.

\end{document}